\documentclass[printer,bibyear]{aa}
\usepackage{graphicx}
\usepackage{txfonts}

\begin{document}
%
   \title{Dynamics of electron beams in the solar corona plasma with density fluctuations}

   \author{Eduard P. Kontar
          }


   \institute{Institute of Theoretical Astrophysics, P.B. 1029 Blindern, 0315 Oslo, Norway\\
              email: {\tt eduard.kontar@astro.uio.no}}

   \date{\today}
\authorrunning{Eduard P. Kontar}
\titlerunning{Dynamics of electron beams in the solar plasma}
 
\abstract{The problem of beam propagation in a plasma with small scale and
low intensity inhomogeneities is investigated. It is shown that the
electron beam propagates in a plasma as a beam-plasma structure
 and is a source of Langmuir waves. The plasma inhomogeneity
changes the spatial distribution of the waves. The spatial
distribution of the waves is fully determined by the distribution of
plasma inhomogeneities. The possible applications to the theory of
radio emission associated with electron beams are
discussed.}
\keywords{Sun -- electron beams -- Langmuir waves -- Type III
bursts}
  \maketitle

\section{Introduction}
One of the challenging problems in the theory of type III bursts,
widely discussed in the literature, is the fine structure of the
bursts. The fine structure is observed in almost all ranges of
frequencies from GHz (\cite{Benz82}; \cite{Benz96}) to a few tens
of kHz in  interplanetary space (\cite{Chaizy95}). Direct
observations of Langmuir waves and energetic electrons show that
Langmuir waves have rather clumpy spatial distribution whereas the
electron stream seems rather continuous (\cite{Lin81};
\cite{Chaizy95}).

There are a few alternative ways to explain the observational
data. The existing theories can be roughly divided into three
groups in accordance with the electron beam density or the energy
of the Langmuir waves. The first group of theories is based on the
assumption that nonlinear instabilities of strong turbulence
theory can suppress quasilinear relaxation (\cite{Papadopoulos74})
and lead to extreme clumpiness of the spatial distribution of
Langmuir waves (\cite{Thejappa98}). However, some observations and
theoretical studies (\cite{Cairns95}) raise doubts as to whether
the Langmuir turbulence level is high enough for strong-turbulence
processes. The second, recently developed group of theories, is
based on the prediction that an electron beam propagates in a
state close to marginal stability, i.e. one where the
fluctuation-dependent growth rate is compensated for by the
damping rate (\cite{Robinson92}; \cite{Robinson93}). In this view,
the growth rate of beam-plasma instability  is perturbed by the
ambient density fluctuations (\cite{Robinson92}). The third, more
traditional group of theories, considers the beam propagation in
the limit of weak turbulence theory (\cite{Ryutov70};
\cite{Takakura76}; \cite{Magelssen77}; \cite{Takakura82};
\cite{Grognard85}). The basic idea is that the electron beam
generates Langmuir waves at the front of the electron stream and
the waves are absorbed at the back of the stream, ensuring
electron propagation over large distances. However, this idea was
not proved for a long time (\cite{Melrose90}).  Recently Mel'nik
has demonstrated analytically (\cite{Melnik95}) that a
mono-energetic beam can propagate as a beam-plasma structure
(BPS). This result has been confirmed numerically
(\cite{Kontar98}) and applied to the theory of type III bursts
(\cite{Melnik99}). The solution obtained (\cite{Melnik00a})
directly resolves Sturrock's dilemma (\cite{Sturrock64}) and may
explain the almost constant speed of type III sources. However,
the influence of plasma inhomogeneity on the dynamics of a BPS has
never been studied although
 the correlation between Langmuir wave clumps and
density fluctuations demonstrates the importance of such
considerations (\cite{Robinson92c}).

The influence of plasma inhomogeneity on Langmuir waves and beam
electrons has been studied from various points of view. An account
of plasma inhomogeneities may explain why accelerated beam
electrons appear in the experiments with quasilinear relaxation of
an electron beam (\cite{Ryutov69}). Relativistic dynamics of an
electron beam with random inhomogeneities, as applied to
laboratory plasmas, was considered in (\cite{Hishikawa76}). It has
been shown (\cite{Muschietti85}) that the solar corona density
fluctuations may be extremely effective in quenching the
beam-plasma instability. Moreover, the isotropic plasma
inhomogeneities may lead to efficient isotropisation of plasma
waves (\cite{Goldman82}) whereas those alongated along the
direction of ambient magnetic field have little influence on the
beam stability. Therefore, the growth rate of beam-plasma
instability was postulated to be very high in the regions of low
amplitude density fluctuations (\cite{Melrose86};
\cite{Melrose87}). Isotropic density fluctuations of ambient
plasma density were also employed to explain low level of Langmuir
waves in microbursts (\cite{Gopalswamy93}).

In this paper the dynamics of a spatially limited electron cloud
is considered in a plasma with small scale density fluctuations.
In the treatment presented here, quasilinear relaxation is a
dominant process and density inhomogeneities are too weak to
suppress the instability. Indeed, observations of interplanetary
scintillations from extragalactic radio sources (\cite{Cronyn72})
lead to an average value of $\Delta n/n$ of the order of $10^{-3}$
(\cite{Smith79}). Nevertheless, the low intensity density
fluctuations lead to significant spatial redistribution of wave
energy. The numerical results obtained demonstrate that electrons
propagate as a continuous stream while the Langmuir waves
generated by the electrons are clumpy. Both electrons and Langmuir
waves propagate in a plasma as a BPS with an almost constant
velocity. However, density fluctuations lead to some energy
losses.

\section{Electron beam and density fluctuations}
The problem of one-dimensional electron beam propagation is
considered in a plasma with density fluctuations. The
one-dimensional character of electron beam propagation is
supported by the 3D numerical solution of the kinetic equations
(\cite{Churaev80}) and additionally by the fact that in the case
of type III bursts electrons propagate along open magnetic field
lines (\cite{Dulk85}).

\subsection{Electron beam}

There is still uncertainty in the literature as to whether
electron beams are strong enough to produce strong turbulence or
whether the beam is so rarified that quasilinear relaxation is
suppressed by damping or scattering. While some observations are
in favor of the strong turbulence regime (\cite{Thejappa98})
others are interpreted as implying marginal stability
(\cite{Cairns95}). Therefore, we consider the intermediate case of
a medium density beam, which is not strong enough to start strong
turbulence processes,
\begin{equation}
W/nT \ll (k\lambda _{D})^2,
 \label{weakt}
\end{equation}
but is dense enough to make quasilinear relaxation a dominant
process. Here $W$ is the energy density of Langmuir waves generated by
the beam, $T$ is the temperature of the  surrounding plasma, $k$
is the wave number, and $\lambda _{D}$ is the electron Debye
length.

The initial value problem is solved with an initially-unstable
electron distribution function, which leads to the formation of a
BPS in the case of homogeneous plasma (\cite{Melnik00a})
\begin{equation}
f(v,x,t=0)=g_0(v)\mbox{exp}(-x^2/d^2),  \label{f_0}
\end{equation}
where
\begin{equation}
g_0(v)=\left\{ \begin{array}{ll} \displaystyle \frac{2n'v}{v_0^2},
&\mbox{$v<v_0$},\\ 0, &\mbox{$v>v_0$}.
\end{array}
\right.
\label{g_0}
\end{equation}
Here $d$ is the characteristic size of the electron cloud and $v_0$ is the
velocity of the electron beam. The initial
spectral energy density of Langmuir waves
\begin{equation}
W(v,x,t=0)\simeq \frac{T}{2\pi ^2\lambda _{D}^2}, \label{W_0}
\end{equation}
is of the thermal level and uniformly distributed in space. The
electron temperature of the corona is taken to be $T=10^{6}$K,
which gives an electron thermal velocity
$v_{Te}=\sqrt{3kT/m}\simeq 6.7\times 10^{8}$cm~s$^{-1}$.

\subsection{Ambient density fluctuations}
Following common practice in the literature on plasma
inhomogeneity Langmuir waves are treated in the approximation of
geometrical optics (the WKB approximation) when the length of a
Langmuir wave is much smaller than the size of the plasma
inhomogeneity (\cite{Vedenov67}; \cite{Ryutov69})
\begin{equation}\label{WKB1}
  \lambda \ll L,
\end{equation}
where
\begin{equation}\label{WKB2}
 L\equiv \left(\frac {1}{\omega _{pe}}\frac{\partial \omega _{pe}}{\partial
 x}\right)^{-1},
\end{equation}
is the scale of ambient plasma density fluctuations, and $\omega
_{pe}$ is the local electron plasma frequency. The plasma
inhomogeneity changes the dispersion properties of Langmuir waves
and if the intensity of density fluctuations is small then the
dispersion relation can be written
\begin{equation}
\label{dispersion} \omega (k,x) =\omega_{pe}\left[1+\frac
12\frac{\Delta n}{n}+\frac{3k^2v_{Te}^2}{2\omega_{pe}^2}\right],
\end{equation}
where $v_{Te}$ is the electron thermal velocity. The intensity of
the density fluctuations should be small (\cite{Coste75})
\begin{equation}\label{delta_n}
\frac{\Delta n}{n} < \frac{3k^2v_{Te}^2}{\omega_{pe}^2},
\end{equation}
to ensure that the corresponding fluctuations of local plasma
frequency are within the thermal width of plasma frequency. Thus, for
the typical parameters of the corona plasma (plasma density
$n=5\times 10^8$cm$^{-3}$ or plasma frequency $f_p=\omega
_{pe}/2\pi \approx 200.73$MHz), and assuming a beam velocity
$v_0=10^{10}$cm~s$^{-1}$, the density
fluctuations are limited to $\Delta n/n<10^{-2}$.

\subsection{Quasilinear equations}

In the case of weak turbulence theory (\ref{weakt}), and under the
conditions of the WKB approximation (\ref{WKB1},\ref{delta_n}), the
evolution of the electron distribution function $f(v,x,t)$ and the
spectral energy density $W(v,x,t)$ are described by the system of
kinetic equations (\cite{Ryutov69})
\begin{equation}
\frac{\partial f}{\partial t}+v \frac{\partial f}{\partial x}=
\frac{4\pi ^2 e^2 }{m^2}\frac{\partial}{\partial v}
\frac{W}{v}\frac{\partial f}{\partial v}, \label{eqk1}
\end{equation}
and
\begin{equation}
  \frac{\partial W}{\partial t}+\frac{\partial
\omega}{\partial k} \frac {\partial W}{\partial x}-\frac{\partial
\omega_{pe}}{\partial x } \frac{\partial W}{\partial k}=\frac{\pi
\omega_{pe}}{n}v^2W\frac{\partial f}{\partial v}, \;
\omega_{pe}=kv, \label{eqk2}
\end{equation}
where $\partial \omega/\partial k=3v_{Te}^2/v$ is the group
velocity of Langmuir waves, and $W(v,x,t)$ plays the same role for
waves as the electron distribution function does for particles.
The system (\ref{eqk1},\ref{eqk2}) describes the resonant
interaction $\omega_{pe}=kv$ of electrons and Langmuir waves. On
the right-hand side of equations (\ref{eqk1},\ref{eqk2}) I
omit the spontaneous terms due to their small magnitude relative to
the induced ones (\cite{Ryutov70}).

The presence of a local plasma frequency gradient leads to two
physical effects on the kinetics of the Langmuir waves
(\ref{eqk2}). Firstly, the characteristic time of the beam-plasma
interaction depends on the local density and therefore the
resonance condition for the plasmons may itself change during the
course of beam propagation. Secondly, the Langmuir wave
propagating in the inhomogeneous plasma experiences a shift of
wavenumber $\Delta k(x)$, due to the variation of the local
refractive index. The second effect has been shown to have the
main impact on Langmuir wave kinetics whereas the first effect can
be neglected (\cite{Coste75}).

Thus, we are confronted with the initial value problem of electron
cloud propagation in a plasma with density fluctuations. The
problem is nonlinear and is characterized by three different time
scales. The fastest process in the system is the quasilinear
relaxation, on the quasilinear timescale $\tau \approx
n/n'\omega_{pe}$. The second timescale is that of processes
connected with plasma inhomogeneity. Thirdly, there is the
timescale of an electron cloud propagation in a plasma that
significantly exceeds all other timescales.

\section{Quasilinear relaxation and plasma inhomogeneity}

The main interaction in the system is beam -- wave interaction
governed by the quasilinear terms on the right hand side of equations
(\ref{eqk1},\ref{eqk2}) . It is well-known that the unstable
electron distribution function (\ref{g_0}) leads to the generation
of plasma waves. The result of quasilinear relaxation for an
electron beam homogeneously distributed in space is a plateau of
the electron distribution function (\cite{Ryutov70})
\begin{eqnarray}
 \displaystyle f(v,t\approx \tau) =\left\{\begin{array}{ll}
              \displaystyle \frac{n'}{v_0},&\mbox{$v<v_0$}\\
              0,&\mbox{$v>v_0$}
              \end{array}
\right. \label{f_relax}
\end{eqnarray}
and the spectral energy density
\begin{eqnarray}
 \displaystyle W(v,t\approx \tau) =
              \displaystyle \frac{mn'}{v_0\omega_{pe}}
\int_0^v\left(1-\frac{v_0}{n'}g_0(v)\right)dv,\;\;v<v_0
\label{w_relax}
\end{eqnarray}
where $g_0(v)$ is the initial distribution function of the beam.

In the case of an inhomogeneous plasma we can also consider
relaxation of a homogeneously distributed beam. Thus, the kinetic
equations (\ref{eqk1},\ref{eqk2}) will take the form
\begin{equation}
\frac{\partial f}{\partial t}= \frac{4\pi ^2 e^2
}{m^2}\frac{\partial}{\partial v} \frac{W}{v}\frac{\partial
f}{\partial v}, \label{k1}
\end{equation}
and
\begin{equation}
\frac{\partial W}{\partial t}+\frac{v^2}{L_0} \frac{\partial
W}{\partial v}=\frac{\pi \omega_{pe}}{n}v^2W\frac{\partial
f}{\partial v}, \;\;\; \omega_{pe}=kv, \label{k2}
\end{equation}
where the transport terms are omitted. Here, the inhomogeneity
scale is also assumed to be constant and equal to $L_0$. It should
be noted that this assumption is physically incorrect. The change
in the spectrum of the Langmuir waves is due solely to the spatial
movement of the waves with the group velocity. However, from a
mathematical point of view, it is well justified as the group
velocity of Langmuir waves is small ($3v_{Te}^2/v\ll v$) and
effects connected with wave transport can be neglected.

Equations (\ref{k1},\ref{k2}) describe two physical effects:
quasilinear relaxation (with characteristic time $\tau $) and the
drift of Langmuir waves in velocity space (the characteristic time
$\tau _2 = |L_0|/v$. Since $\tau _2\gg \tau$ the influence of
plasma inhomogeneity can be considered as the evolution of the
final stage of quasilinear relaxation. Two possible cases of
plasma density change are considered: plasma density decreasing
($L_0<0$) with distance and plasma density increasing ($L_0>0$)
with distance.

\subsection{Plasma density decreasing with distance}

In this case $L_0$ is negative. After the time of
quasilinear relaxation, a plateau is established in the electron
distribution function and a high level of Langmuir waves is
generated. Since the quasilinear processes are fast we have a
plateau at every  moment of time
\begin{eqnarray}
 \displaystyle f(v,t\approx \tau) =\left\{\begin{array}{ll}
              \displaystyle \frac{n'}{v_0},&\mbox{$v<v_0$}\\
              0,&\mbox{$v>v_0$}
              \end{array}
\right. \label{relax2}
\end{eqnarray}
The wave spectrum is changing with time, and from the fact that we
have a plateau at every moment equation (\ref{k2}) can be reduced to
\begin{equation}
\frac{\partial W}{\partial t}-\frac {v^2} {|L_0|} \frac{\partial
W}{\partial v}= 0 \label{W}
\end{equation}
The role of initial wave distribution is played by the spectral
energy density generated during the relaxation stage
(\ref{w_relax}). Integrating equation (\ref{W}) we obtain the
solution for $t\gg \tau$
\begin{eqnarray}
\label{w1} W(v,t)=\frac{m}{\omega _{pe}}(1/v-t/|L_0|)^{-3}
\;\;\;\;\;\;\;\;\;\;\;\;\;\;\;\;\;\;\;\;\;\;\;\;\;\;\;\;\;\;\;\;\;\cr
\;\;\;
\times\int_0^{1/(1/v-t/|L_0|)}\left[\frac{n'}{v_0}-g_0(v)\right]dv,\;\;
              v<u(t)
\end{eqnarray}
where
\begin{equation}\label{u}
  u(t)=\frac{v_0}{1+v_0t/|L_0|}
\end{equation}
is the maximum velocity of the Langmuir waves. Note, that the electron
distribution function is constant and presents a plateau
(\ref{relax2}).

\begin{figure}
\includegraphics[width=90mm]{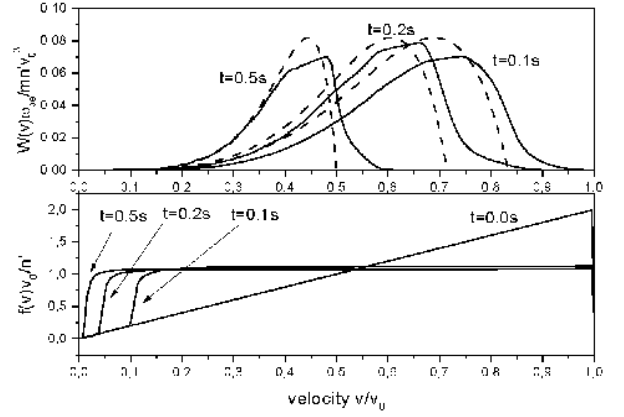}
 \caption{The electron distribution function $f(v,t)$ and the spectral
 energy density of Langmuir waves $W(v,t)$ at various times, for the case
where the
 plasma density decreases with distance, $L_0=-5\times 10^9$cm. Numerical
 solution of kinetic equations (\ref{k1},\ref{k2}) $n'=100$cm$^{-3}$,
 $v_0=1.0\times 10^{10}$cm~s$^{-1}$.}
 \label{fig1}
\end{figure}

The numerical solution of  equations (\ref{k1},\ref{k2}) with
the initial electron distribution function (\ref{g_0}) is
presented in fig. \ref{fig1}. Comparing the numerical
results and the simplified solution (\ref{w1}) we see a good
agreement (see fig. \ref{fig1}). The plateau for a wide range of
velocities is formed after a short time, $t=0.1$s, and it remains
almost unchanged up to the end of the calculation. For the time
$t>0.1$s, the drift of the Langmuir wave spectrum toward smaller phase
velocities becomes observable. At $t=0.5$s, the maximum phase
velocity is half of the initial beam velocity.

\subsection{Plasma density increasing with distance}

An increasing plasma density leads to a shift toward
larger phase velocities. For $v>v_0$ we have a negative derivative
at the edge of the electron distribution function, and electrons
absorb waves with the corresponding phase velocities. Absorption
of waves then leads to acceleration of particles. This process
continues until all the waves generated during the beam relaxation are
absorbed by the electrons.

In the case of increasing density we are unable to find an exact
solution, but we can find the solution for $t\rightarrow \infty$. Using
conservation of energy (\cite{Ryutov69})
\begin{eqnarray}\label{e_cons}
  \omega _{pe}\int_0^{u(t)}\frac{W(v,t)}{v^2}dv +
  \int_0^{u(t)}\frac{mn'}{2u(t)}v^2dv \cr
  =\frac m2 \int_0^{v_0}g_0(v)v^2dv
\end{eqnarray}
and that the fact $W(v,t)=0$ at $t\rightarrow \infty$ we can find
the maximum velocity for the initial distribution function
(\ref{g_0})
\begin{equation}\label{u_max}
  u(t\rightarrow \infty) =\sqrt{3/2}v_0
\end{equation}

\begin{figure}
\includegraphics[width=90mm]{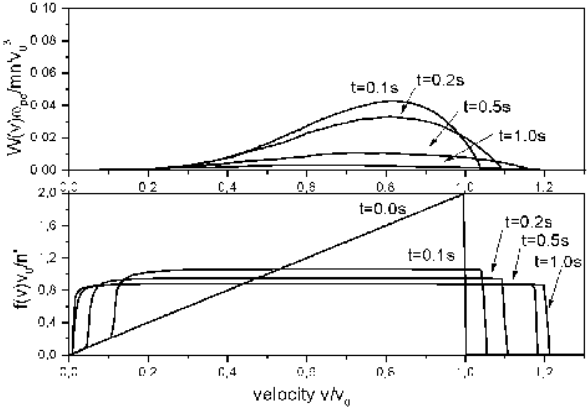}
 \caption{The electron distribution function $f(v,t)$ and the spectral
 energy density of Langmuir waves $W(v,t)$ at various times for the case
where the plasma density
 increases with distance, $L_0=5\times 10^9$cm. Numerical
 solution of kinetic equations (\ref{k1},\ref{k2}) $n'=100$cm$^{-3}$,
 $v_0=1.0\times 10^{10}$cm~s$^{-1}$.}
 \label{fig2}
\end{figure}

As predicted, the numerical solution tends to the maximum velocity
$\approx 1.22v_0$ (fig. \ref{fig2}). As in the previous case, 
at $t=0.1$s we have the result of quasilinear relaxation - a plateau
in the electron distribution function and a high level of Langmuir
waves. For times $t>0.1$s, the drift of Langmuir waves and
consequent acceleration of electrons is observable. At $t=1$s
almost all plasma waves are observed near the leading edge of the
plateau and the maximum plateau velocity is close to the value
given by (\ref{u_max}).

\section{Propagation of an electron cloud}

In this section the numerical results of the evolution of the
electron beam in the plasma with density fluctuations are
presented. We begin with the case where the ambient density
fluctuations in the plasma are periodic and sine-like. The
dependency of plasma density on distance is
\begin{equation}\label{sin}
  n(x)= n_0(1+\alpha \mbox{sin}(x/\Delta x))
\end{equation}
where $\Delta x$ defines the period of the density fluctuations
and $\alpha n_0$ is the amplitude of the density irregularities.
The background plasma density is taken as a typical value for the
starting frequencies of type III bursts $n_0=5\times
10^8$~cm$^{-3}$, corresponding to a local plasma frequency
$f_p=\omega _{pe}/2\pi=200.73$MHz. As noted, small-intensity
density fluctuation are considered, i.e. the local plasma
frequency change due to the inhomogeneity is less than the thermal
width of the plasma frequency (\ref{delta_n}). The value $\alpha$
is taken to be $10^{-3}$, which is considered to be a typical
value for solar coronal observations (\cite{Cronyn72};
\cite{Smith79}). The spatial period of the plasma fluctuations
$\Delta x=d/12$ is taken to be less than the initial size of the
electron cloud. Thus, we have regions of size $\pi d/12\approx
0.26d$ with positive and negative density gradients. Recently, it
has been shown that an electron beam can propagate in a
homogeneous plasma as a BPS (\cite{Melnik99}; \cite{Melnik00};
\cite{Melnik00a}). Therefore, it is important to consider the
dynamics of the electron beams at distances greatly exceeding the
size of the electron cloud.

\subsection{Initial evolution of the electron beam and formation
of a BPS}

At the initial time $t=0$ we have an electron distribution
function which is unstable. Due to
fast quasilinear relaxation, electrons form a plateau in the
electron distribution function and generate a high level of plasma
waves. At time $t=0.1$s, the typical result of
quasilinear relaxation is observed. The electron distribution
function and the spectral energy density evolve in accordance with
the gas-dynamic solution (\cite{Melnik95}; \cite{Melnik00a})
\begin{eqnarray}
 \displaystyle f(v,x,t) =\left\{\begin{array}{ll}
              \displaystyle \frac{n'}{v_0}\mbox{exp}\left(-\frac{(x-v_0t/2)^2}{d^2}\right),
              &\mbox{$v<v_0$}\\
              0,&\mbox{$v>v_0$}
              \end{array}
\right. \label{f_hom}
\end{eqnarray}
\begin{eqnarray}
\label{w_hom} W(v,x,t)=\frac{mn'}{v_0\omega _{pe}}v^4
              \left[1-\frac{v}{v_0}\right]
              \;\;\;\;\;\;\;\;\;\;\;\;\;\;\;\;\;\;\;\;\;\;\;\;\;\;\;\cr
              \;\;\;\;\;\;\;\;\;\;\;\;\;\;\;\;\;\;\;\;\;\;
              \times\mbox{exp}\left(-\frac{(x-v_0t/2)^2}{d^2}\right), \;\;v<v_0
\end{eqnarray}
At this stage the influence of the plasma inhomogeneity is not
observable.

The numerical solution of the kinetic equations and the
gas-dynamic solution show that electrons propagate in a plasma
accompanied by a high level of plasma waves. Since the plasma
waves exist at a given point for some time, while the structure
passes this point, the spectrum of the waves should change due to
the wave movement. To understand the physics of the process we
consider the evolution of the electron distribution function and
the spectral energy density of Langmuir waves at a given point.

\subsection{The electron distribution function and the spectral energy
density of plasma waves}

At every spatial point we observe two physical processes. The
first process is connected with the spatial movement of a BPS, as
would be the case for a homogeneous plasma (\cite{Melnik00};
\cite{Melnik00a}). The second process is the influence of plasma
inhomogeneity on the Langmuir waves. Depending on the
 sign of the density gradient, the  Langmuir wave spectrum
takes on a different form.

\begin{figure}
\includegraphics[width=90mm]{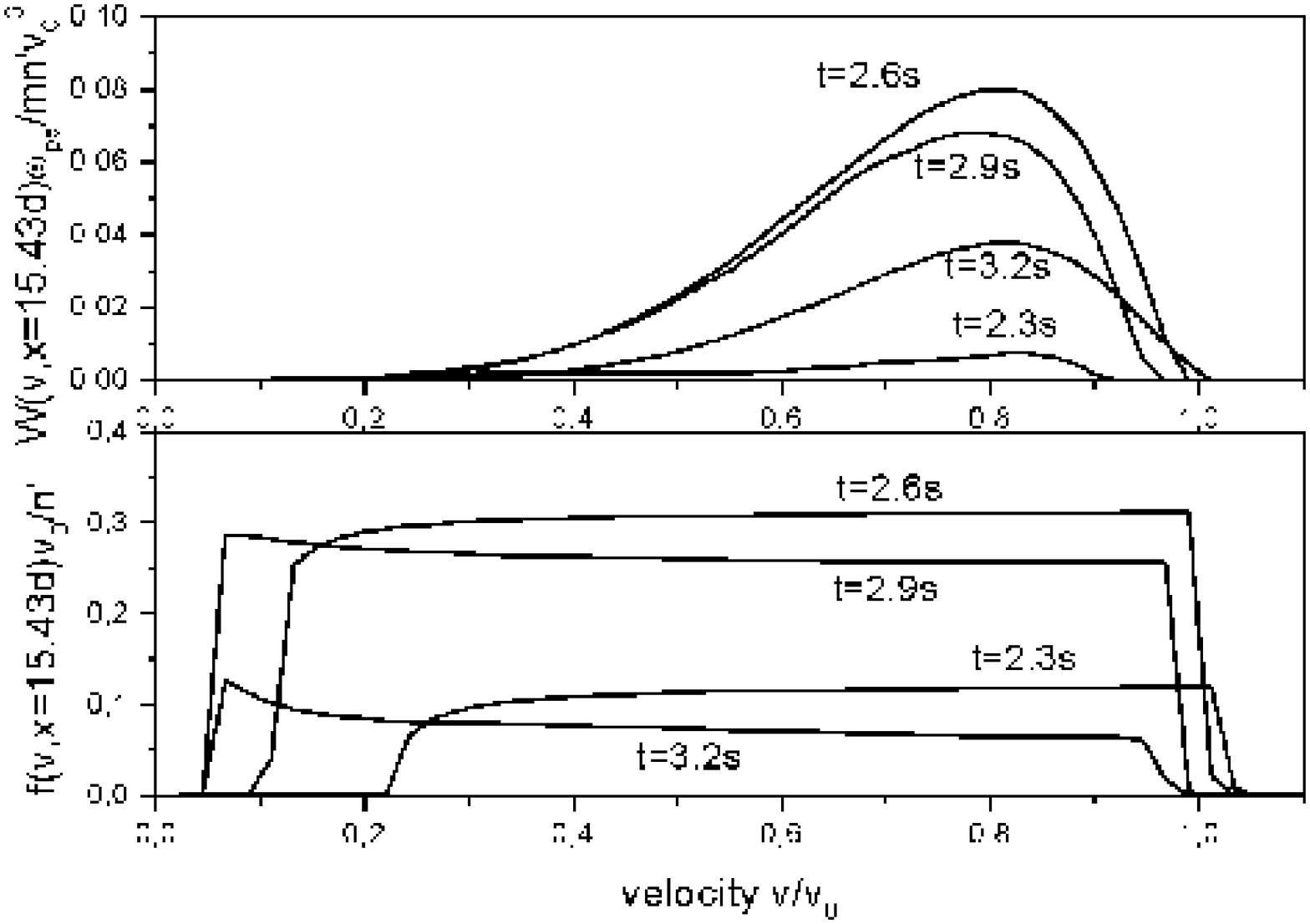}
\includegraphics[width=90mm]{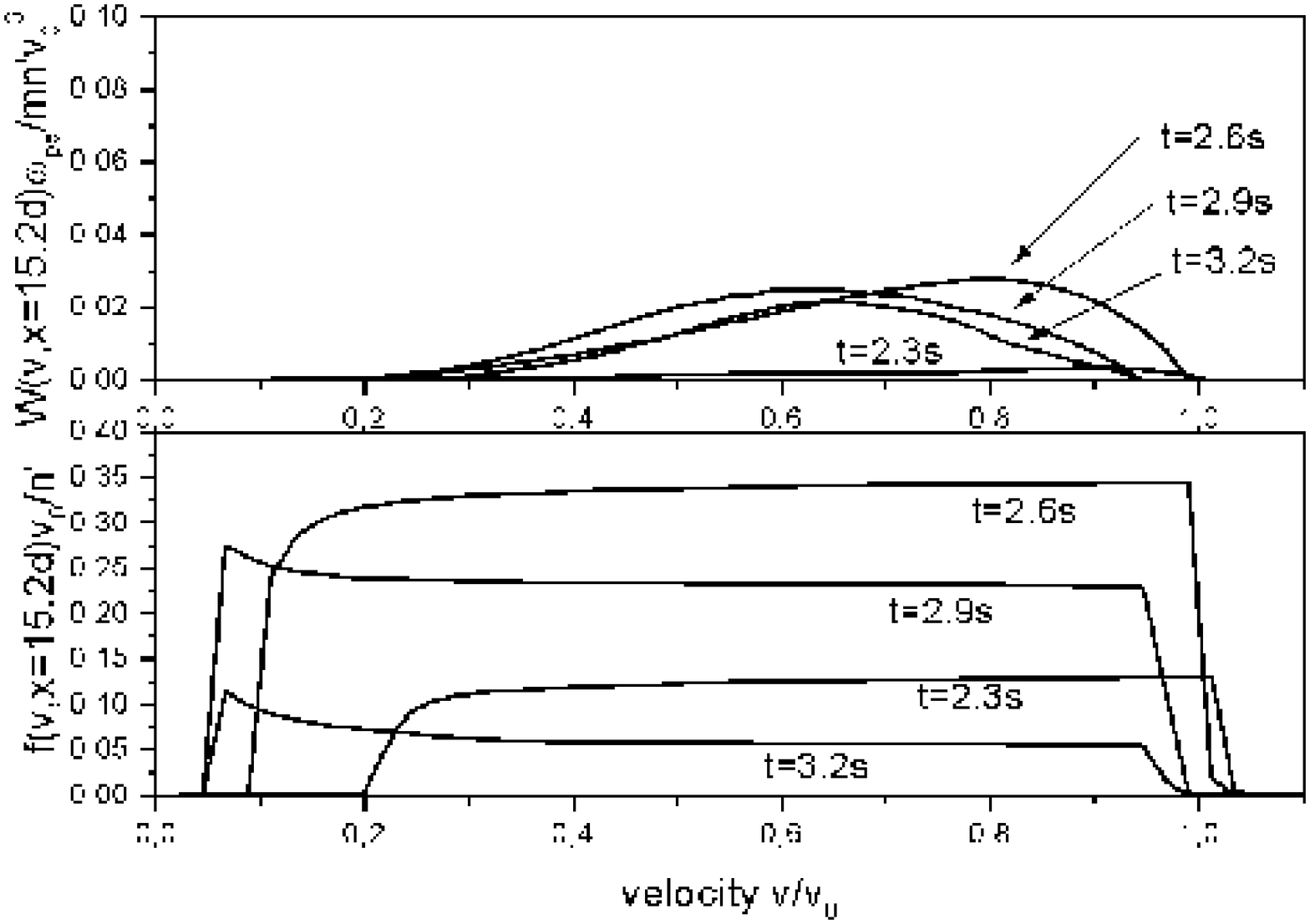}
 \caption{The electron distribution function $f(v,x,t)$ and the spectral
 energy density of Langmuir waves $W(v,x,t)$ at $x=15.2d$ and at $x=15.47d$. Numerical
 solution of kinetic equations for a plasma with sine-like density
 fluctuations (\ref{sin}) $n'=100$cm$^{-3}$, $v_0=1.0\times 10^{10}$cm~s$^{-1}$.}
 \label{fig3}
\end{figure}

Consider the time evolution of the electron distribution
function and the spectral energy density of Langmuir waves at two
close points $x=15.2d$ and $x=15.47d$ (see fig. \ref{fig3}). The first
point is chosen in the region with increasing density and the
second in the region where the density decreases with distance. The
first particles arrive to these points at approximately  $t\sim
1.9$s. The arriving electrons form a plateau in the electron
distribution function and generate a high level of plasma waves for
the time of quasilinear relaxation $\tau \approx  0.01$ s.

The movement of the particles leads to the growth of the plateau
height at the front of the structure for $1.9\mbox{ s}<t<2.7\mbox{
s}$. Due to the fact that at the front of the structure electrons
come with a positive derivative, $\partial f/\partial v
>0$, the level of plasma waves also increases.
When the peak of the plateau height is reached at $t\approx 2.7$s
the reverse process takes place. The plateau height decreases and
the arriving electrons have a negative derivative $\partial
f/\partial v <0$ that leads to absorption of waves. The growth
and decrease of the plateau height and the level of plasma waves
is typical for a homogeneous plasma (\cite{Melnik00a}). However,
while the structure passes a given point the spectrum of Langmuir
waves experiences the change. This change depends on the sign of
the plasma density gradient. In the region with decreasing
density ($x=15.47d$) the Langmuir waves have a negative shift in
velocity space while the growing plasma density ($x=15.2d$)
supplies a positive shift in phase velocity of the plasma waves. At
the point with the positive gradient, the Langmuir waves shifted in
phase velocity space are effectively absorbed by the electrons
while the negative plasma gradient does not lead to the absorption
of waves. This behavior results in different levels of plasma waves at
two very close points with the opposite density-gradient sign.

Figure~\ref{fig3} demonstrates the existence of accelerated
electrons with $v>v_0$. These electrons are accelerated by
Langmuir waves in the regions with positive plasma-density
gradient. Electrons with velocity larger than the initial
beam velocity have been observed in laboratory plasma
experiments. This effect was also considered from an analytical
standpoint by
(\cite{Ryutov69}) in application to laboratory plasmas.

\subsection{Dynamics of electrons and accompanying Langmuir waves}

The processes of wave generation at the front and absorption at
the back take place at every spatial point and therefore the
structure can travel over large distances, being the source of
plasma waves (\cite{Kontar98}; \cite{Melnik99}; \cite{Melnik00a}).

\begin{figure}
\includegraphics[width=90mm]{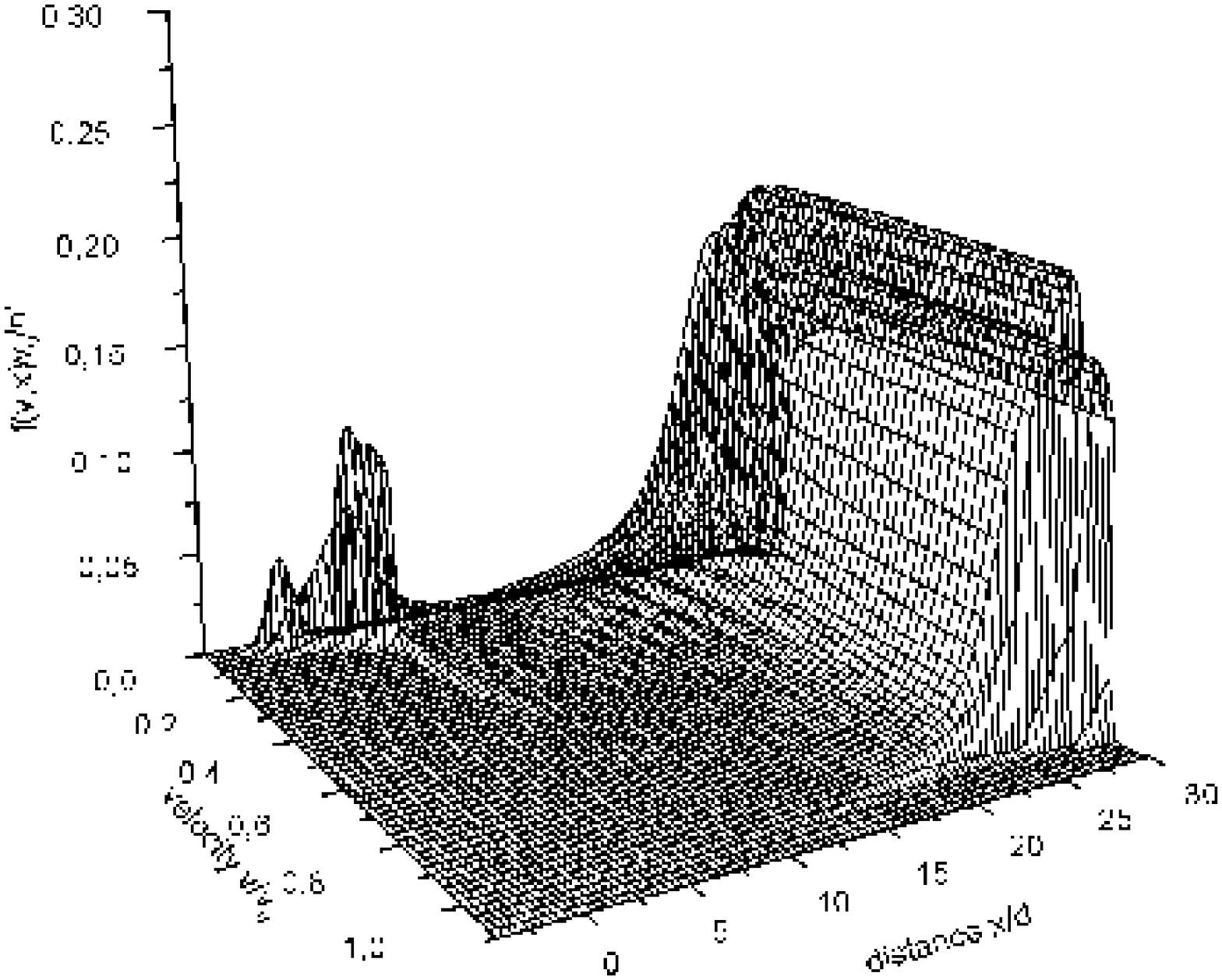}
\includegraphics[width=90mm]{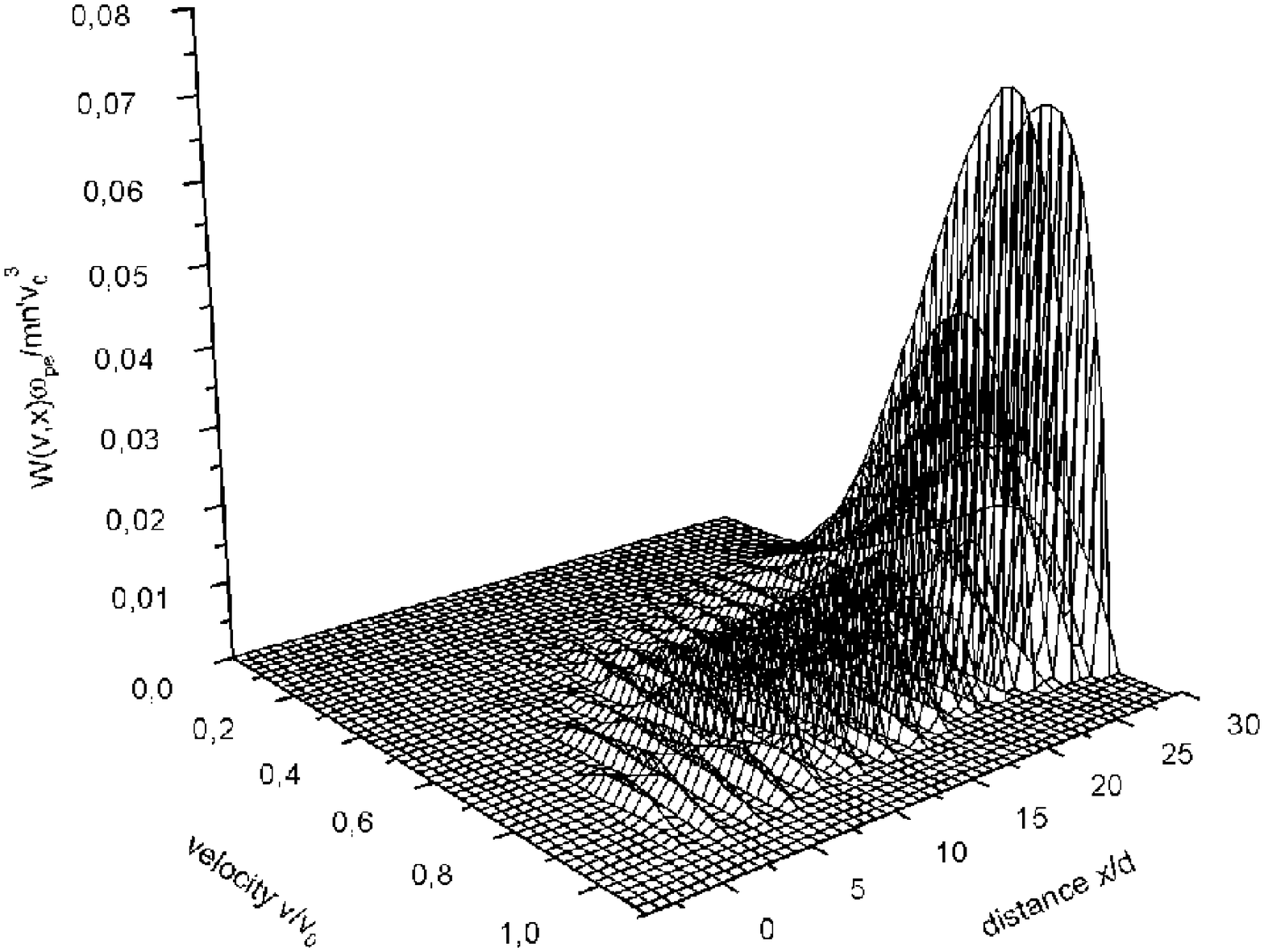}
 \caption{The electron distribution function $f(v,x,t)$ and the spectral
 energy density of Langmuir waves $W(v,x,t)$ at $t=5.0$s. Numerical
 solution of the kinetic equations with sine-like density
 fluctuations (\ref{sin}) $n'=100$cm$^{-3}$,
 $v_0=1.0\times 10^{10}$cm~s$^{-1}$.}
 \label{fig4}
\end{figure}

At time $t=5.0$s, electrons accompanied by Langmuir
waves have passed over a large distance but the general physical
picture remains the same (fig. \ref{fig4}). Generally, electrons
and Langmuir waves propagate as a BPS. At every spatial point
electrons form a plateau at the electron distribution function and
we have a high level of plasma waves. The electron cloud has a
maximum of the electron density at $x=27d$. Plasma waves are also
concentrated in this region and the maximum of Langmuir wave
density is located at the maximum of electron density $x=27d$. The
spectrum of Langmuir waves has a maximum close to $v\approx
0.8v_0$. The spatial profile, averaged over the plasma inhomogeneity
period, is close to the result obtained for a homogeneous plasma.

\begin{figure}
\includegraphics[width=90mm]{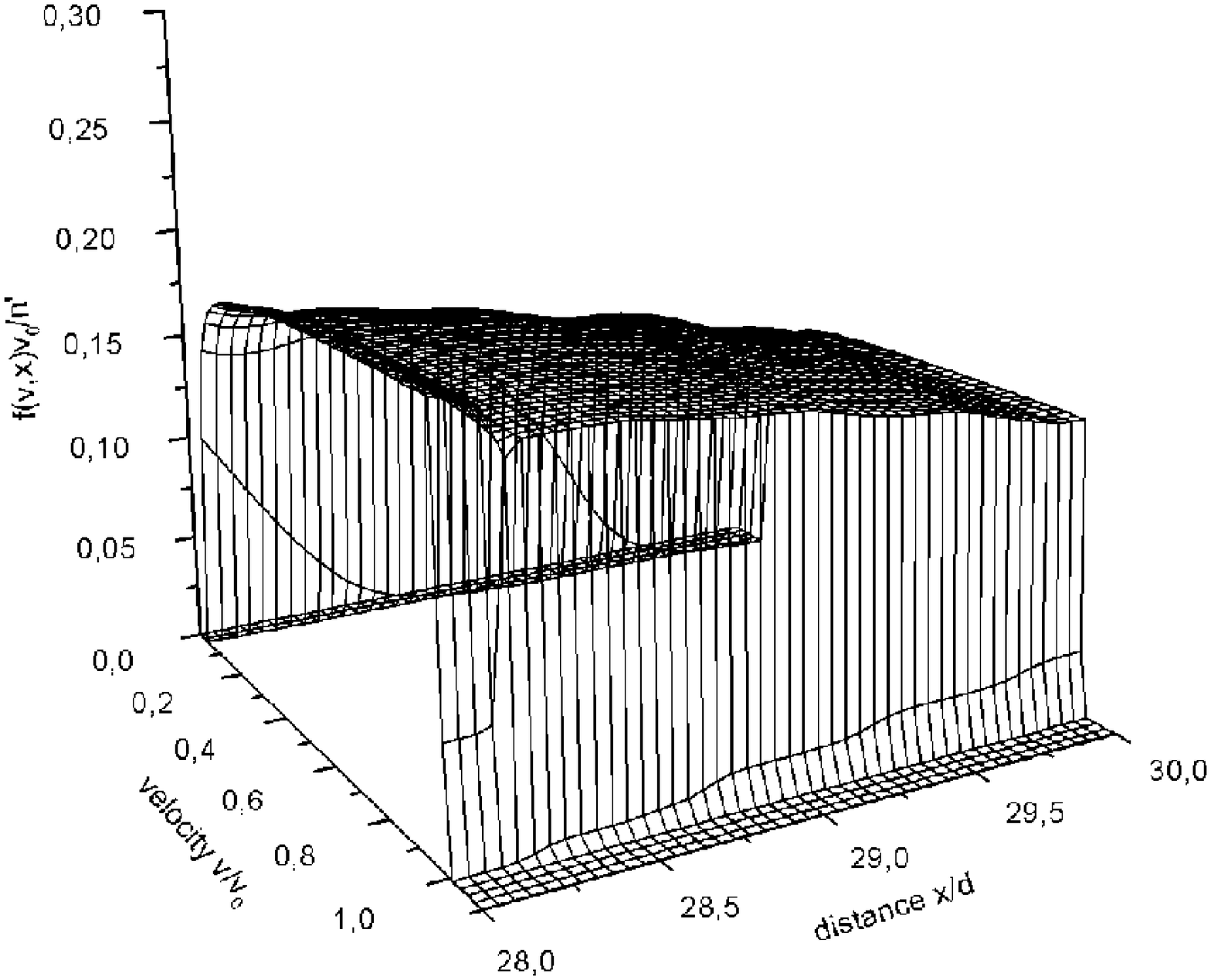}
\includegraphics[width=90mm]{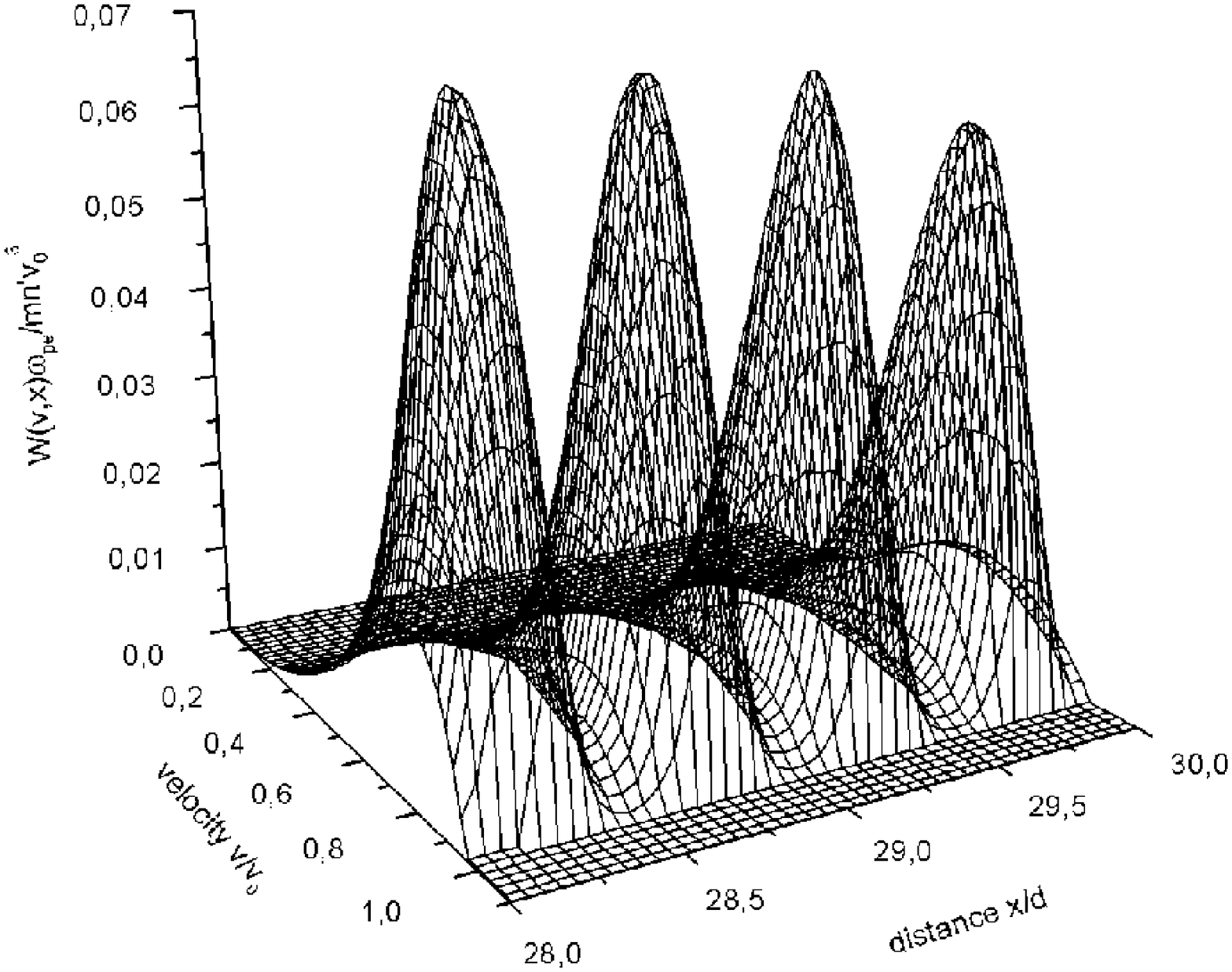}
 \caption{Detailed picture of the electron distribution function $f(v,x,t)$ and the spectral
 energy density of Langmuir waves $W(v,x,t)$ at $t=5.0$s. Numerical
 solution of kinetic equations with sine-like density
 fluctuations (\ref{sin}) $n'=100$cm$^{-3}$,
 $v_0=1.0\times 10^{10}$cm~s$^{-1}$.}
 \label{fig5}
\end{figure}

However, the spatial profile of Langmuir waves has a fine
structure that can be seen in fig. \ref{fig5}. The Langmuir waves
are grouped into clumps (the regions with high level of plasma
waves, following the terminology of (\cite{Smith79})). The size of
a clump is determined by the spatial size of the density
fluctuations and is equal to half of the density fluctuation
period $\pi d/12\approx 0.26 d$. The maxima of Langmuir wave
density are located in regions of negative plasma-density gradient
and the regions with low levels of Langmuir turbulence are where
the density gradient is positive.

The other interesting result is that while the Langmuir wave
distribution is determined by the irregularities of ambient plasma
the electron distribution function is a smooth function of
distance (see fig. \ref{fig5}). The electrons in the structure
propagate as a continuous stream, being slightly perturbed by the
density fluctuations. The influence of plasma inhomogeneity on
electron distribution is observed in the appearance of accelerated
particles with $v>v_0$ and the fact that the maximum plateau
velocity is slightly decreasing with time during the course of
beam-plasma passing a given point (fig. \ref{fig3}). The
accelerated electrons tend to accumulate at the front of the
structure and the electrons decelerated concentrate at the back of
the structure.

\subsection{The energy distribution of waves}

\begin{figure}
\includegraphics[width=90mm]{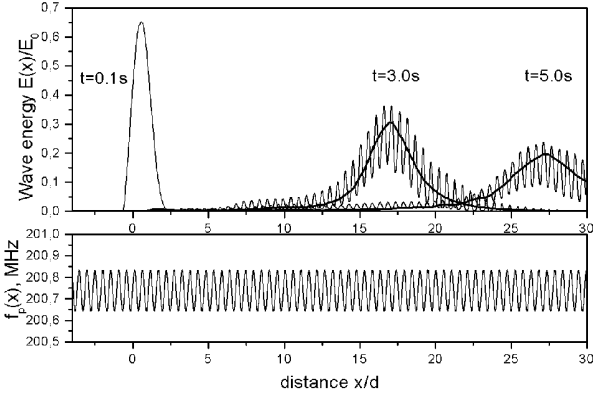}
 \caption{The energy density of plasma waves $E(x)$ at various
 times and the local plasma frequency $f_p(x)$ (\ref{sin}) as a
 function of distance at various times. The bold line shows the numerical
 solution for homogeneous plasma. Numerical solution of
 kinetic equations  $n'=100$cm$^{-3}$, $v_0=1.0\times 10^{10}$cm~s$^{-1}$.}
 \label{e_sin}
\end{figure}

The energy distribution of waves
\begin{equation}\label{we}
  E_w(x,t)=\int_0^{\infty}Wdk=\omega_{pe}\int_0^{v_0}\frac{W(v,x,t)}{v^2}dv
\end{equation}
is presented in fig. \ref{e_sin}, where $E_0=mn'v_0^2/4$ is the
initial beam energy. The energy distribution explicitly shows the
correlation between the plasma and wave energy-density
fluctuations. The regions of decreasing plasma density have higher
levels of Langmuir turbulence than the corresponding regions with
increasing plasma density. The energy distribution of waves
appears to be modulated by the ambient plasma density fluctuations. On the
other hand, the wave energy density distribution averaged over the
period of density fluctuations has a spatial profile close to
that in a homogeneous plasma. The maximum of wave energy together
with the maximum of electron density propagate with the constant
velocity $\approx 0.5v_0$.

The other physical effect that should be noted is the energy
losses by the structure in the form of Langmuir waves. In
figs.~\ref{fig4},\ref{e_sin} we see that there is a small but
non-zero level of plasma waves behind the beam-plasma structure.
These waves are also concentrated into clumps in the regions where
the plasma gradient is negative. To explain why the structure
leaves the plasma waves we note the negative shift in phase
velocity of Langmuir waves in the regions with a decreasing
density. Due to this shift we have more waves with low phase
velocity than the electrons are able to absorb at the back. As a
result the low velocity waves form a "trace" of the structure
(\cite{Kontar01}).

As was discussed previously, the quasilinear time is small but
finite value. Therefore, the BPS experiences spatial expansion
(\cite{Kontar98}). The initial width-at-half-height of the
structure is less than $2d$ whereas the spatial width of the
structure at $t=5.0$s is about $5d$. Most of the energy and the
majority of particles are concentrated within the width of the
structure. Since the quasilinear time depends on the beam density,
the quasilinear time for the particles far from the center of the
structure is much larger than for the structure electrons. In
these regions we can observe the situation where the influence of
the plasma inhomogeneity is comparable with the quasilinear time.
Indeed, in the tail of the structure we have regions with zero
level of waves (where the Langmuir waves are absorbed by electrons
when the plasma density increases) and regions with Langmuir waves
(where plasma density is decreasing).

\subsection{Pseudo-random fluctuations of density}

\begin{figure}
\includegraphics[width=90mm]{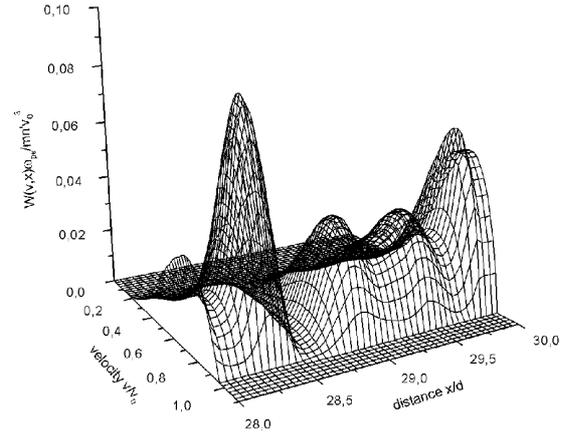}
\includegraphics[width=90mm]{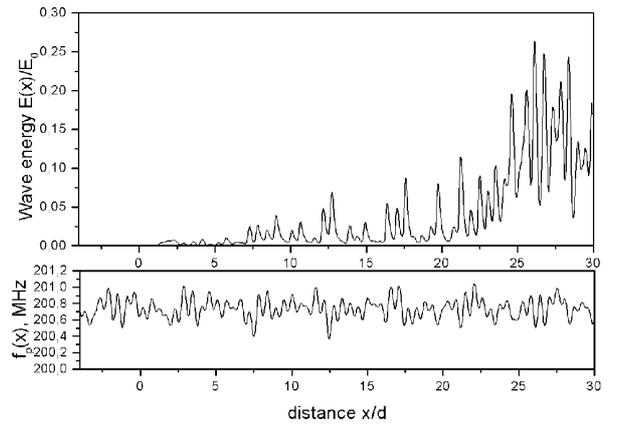}
 \caption{The spectral energy density of Langmuir waves
$W(v,x,t)$, the energy density of plasma waves $E(x)$ at $t=5.0$s,
and the local plasma frequency $f_p(x)$ (\ref{random}) as
functions of distance. Numerical solution of kinetic equations with
random density fluctuations (\ref{random}) $n'=100$cm$^{-3}$,
$v_0=1.0\times 10^{10}$cm~s$^{-1}$.}
 \label{fig7}
\end{figure}

There is special interest in the case where the density fluctuations are
random, which looks like the case for a solar coronal plasma. A
pseudo-random distribution of density fluctuations can be easily
built by summing $N$ sine-like perturbations with random amplitude,
phase, and period
\begin{equation}\label{random}
  n(x)= n_0(1+ \sum \limits_{i=1}^{N}\alpha _i \mbox{sin}(x/\Delta x_i+\varphi _i))
\end{equation}
where $n_0\alpha_i$, $\Delta x_i$, $\varphi _i$ are the amplitude,
period and phase of a given sine-like density oscillations
respectively. The values are chosen in the range to ensure the
applicability of the kinetic equations. Thus, $0<\alpha_i\leq
0.001$, $d/2\leq \Delta x_i\leq d/12$, $0<\varphi _i\leq2\pi$,
$N=10$ are taken for the numerical calculations. The resulting
density profile can be seen in fig.~\ref{fig7}.

The spatial distribution of waves has now more complex structure
(fig. \ref{fig7}). However, all the main results obtained for
sine-like density fluctuations are also observed for
pseudo-random density fluctuations (\ref{random}). Firstly, the
electron stream propagates in a plasma as a BPS. Secondly,
observing the energy density profile of Langmuir waves one can see
the clumps of Langmuir waves. The size of the clumps is determined
by the size of the regions with negative density gradient. The
electron distribution function of beam electrons remains smooth as
in the previous case with sine-like density oscillations.

The dependence of wave energy
density on the amplitude of the density fluctuations is of special
interest. From equation
(\ref{eqk2}) it follows that a Langmuir wave propagating with the
group velocity $v_{gr}=3v_{Te}^2/v$ over the distance $\Delta l$
experiences a shift of phase velocity
\begin{equation}\label{delta_v}
  \Delta v \approx \frac{v^2}{Lv_{gr}}\Delta l.
\end{equation}
Using the density profile (\ref{sin}) and estimating  $L\approx
\Delta l/\alpha$ one derives that
\begin{equation}\label{delta_v2}
  \frac{\Delta v}{v} \approx \frac{1}{3}\left(\frac{v}{v_{Te}}\right)^2\alpha
\end{equation}
where we obtain, for our parameters, a phase velocity shift $\leq
0.1 v$. Expression (\ref{delta_v2}) also demonstrates that the
shift of the wave phase velocity linearly depends on the amplitude
of the density fluctuations. Therefore, in the case with an
arbitrary amplitude of the density fluctuations, the higher the
amplitude of the plasma inhomogeneity the larger the variations of
the wave energy distribution. This tendency can be observed in
fig. \ref{fig7}.

\section{Main results and discussion}

From a physical point of view it is interesting to consider the
physical processes which lead to the reported results. As we see,
the main physical effect, which leads to a complex spatial
distribution of waves, is the shift of the phase velocity $\Delta
v$ due to the wave movement. The growth rate of beam-plasma
instability
\begin{equation}
 \gamma (x)=\frac{\pi
\omega_{pe}}{n}v^2\frac{\partial f}{\partial v}, \label{increment}
\end{equation}
also depends on distance. However, this dependency of the
instability increment on local plasma density is negligible. At
every spatial point we have a plateau with $\partial f/\partial
v\approx 0$ and the value of $\partial f/\partial v$ is determined
by the dynamics of a BPS not by the local plasma density.
Therefore, the shift in phase velocity dominates the effect of
instability increment dependency on distance. Indeed, if we
manually exclude the terms connected with the velocity shift of
the Langmuir waves, the spatial profile of waves will become
smooth and the solution will be close to that obtained in the case
of uniform plasma. This result agrees well with the qualitative
results of (\cite{Coste75}).

For application to the theory of type III bursts, special
interest is presented by a combination of the two main properties of the
solutions.

On one hand, the electron beam can propagate in a plasma over
large distances, and is a source of a high level of Langmuir waves. A
portion of these  Langmuir waves can easily be transformed into observable
radio emission via nonlinear plasma processes (\cite{Ginzburg58}).
At a scale much greater than the size of the beam, electrons and
Langmuir waves propagate as a BPS that may be the source of type
III bursts. The BPS propagates in inhomogeneous plasma with
velocity $\approx v_0/2$ that can explain the almost-constant speed
of the type III source. The finite size of the structure, the
spatial expansion of the structure, and conservation of the
particle number, are promising results for the theory of type III
bursts.

On the other hand, plasma inhomogeneity brings additional
results. The spatial distribution of Langmuir wave energy is
extremely spikey and the distribution of waves is fully determined
by the fluctuations of the ambient plasma density. This fact is in
good agreement with satellite observations
(\cite{Robinson92c}). Moreover, following the plasma emission
model,  one obtains the fine structure of the radio emission.

At  distances about $1$AU the quasilinear time might have a large
value and the characteristic time of a wave velocity shift could
be comparable to the quasilinear time. Therefore, the region of
growing plasma density may lead to the suppression of
quasilinear relaxation, whereas, in regions with a decreasing
density, relaxation is found. Thus the Langmuir waves might be
generated in only those spatial regions where the plasma gradient
is less than or equal to zero. Indeed, in the tails of a beam-plasma
structure the electron beam density is low and Langmuir waves are
only observed in certain regions with non-positive density
gradient.

\section{Summary}

In this paper the dynamics of a spatially bounded electron beam
has been considered. Generally, the solution of the kinetic
equations present a BPS. The structure moves with approximately
constant velocity $\approx v_0/2$ and tends to conserve the number
of particles. As in case of uniform plasma, electrons form a
plateau and generate a high level of plasma waves at every spatial
point.

However, small-scale inhomogeneity in the ambient plasma leads to
significant changes in the spatial distribution of Langmuir
waves. It is found that low intensity oscillations perturb the
spatial distribution of Langmuir waves whereas the electron
distribution function remains a smooth function of distance. The
other interesting fact is that the distribution of waves is
determined by the distribution of plasma inhomogeneities. The
energy density of Langmuir waves has maxima and minima in the
regions with positive and negative density gradient respectively.

Nevertheless, more detailed analysis is needed. One needs to
include radio emission processes in order to calculate the
observational consequences of the model in greater detail. The other challenge is
the detailed comparison of such numerical results with
satellite observations near the Earth's orbit.

\begin{acknowledgements}
Author is extremely thankful to C. Rosenthal for his kind help in
the manuscript preparation.
\end{acknowledgements}

\end{document}